\documentclass[sigconf]{acmart}

\acmConference[MSR 2022]{MSR '22: Proceedings of the 19th International Conference on Mining Software Repositories}{May 23–24, 2022}{Pittsburgh, PA, USA}

\usepackage{multirow}
\usepackage{url}

\newcommand{\numberofprojects}{44 }
\newcommand{\numberofIssues}{508,963 }
\newcommand{\numberofUsers}{208,811 }

\AtBeginDocument{%
  \providecommand\BibTeX{{%
    \normalfont B\kern-0.5em{\scshape i\kern-0.25em b}\kern-0.8em\TeX}}}

\begin{document}
\title{A Versatile Dataset of Agile Open Source Software Projects}

\author{Vali Tawosi, Afnan Al-Subaihin, Rebecca Moussa, Federica Sarro}
\email{{vali.tawosi, a.alsubaihin, rebecca.moussa.18, f.sarro}@ucl.ac.uk}
\affiliation{%
  \institution{University College London}
  \city{London}
  \country{UK}
}

\renewcommand{\shortauthors}{Tawosi, Al-Subaihin, Moussa, and Sarro}

\begin{abstract}
  Agile software development is nowadays a widely adopted practise in both open-source and industrial software projects. Agile teams typically heavily rely on issue management tools to document new issues and keep track of outstanding ones, in addition to storing their technical details, effort estimates, assignment to developers, and more. Previous work utilised the historical information stored in issue management systems for various purposes; however, when researchers make their empirical data public, it is usually relevant solely to the study's objective. In this paper, we present a more holistic and versatile dataset containing a wealth of information on more than 500,000 issues from \numberofprojects open-source Agile software, making it well-suited to several research avenues, and cross-analyses therein, including effort estimation, issue prioritization, issue assignment and many more. We make this data publicly available on GitHub to facilitate ease of use, maintenance, and extensibility.
\end{abstract}



\maketitle

\section{Introduction}
	The early 2000s has witnessed a surge of the adoption of Agile Software Development alongside the release of the \textit{Agile Software Development Manifesto} in 2001 \cite{fowler2001agile}. Agile techniques boast a faster response to unanticipated alterations that can arise during development such as changes in user requirements, development environments and delivery deadlines; typically contrasted with traditional `plan-based' project development, which operates under the assumption that software is specifiable and predictable \cite{Dyba}. Agile Software Development is currently among the most common software development methods in project management \cite{PMI}.
	
	Managing agile software development is commonly aided by an issue tracking tool, which allows agile teams to log and organize outstanding development tasks (e.g.,  bug fixes, functional and non-functional enhancements), in addition to hosting meta-data related to these tasks. Issue tracking tools, such as Jira \cite{jiraweb}, provide a trove of historical information regarding project evolution that promise great value for Empirical Software Engineering research. Such data has been employed to address many software engineering problems such as effort estimation \cite{deep2018, tawosi2022saner}, task prioritization \cite{umer2019cnn,Gavidia2021,huang2021characterizing}, task assignment \cite{mani2019deeptriage}, task description enhancement \cite{chaparro2017detecting}, iteration planing \cite{choetkiertikul2017predicting} and exploring social and human aspects \cite{Ortu2015,ortu2016emotional,valdez2020sentiment, zhang2020sentiment}. 
	However, the data made available by previous empirical studies is usually mainly relevant solely to the study's objective. 
	Therefore, we aim at paving the way for a more holistic and versatile dataset containing a wealth of information on open-source software projects, which can serve as a single source for many possible research avenues, and enable novel investigations on the inter-play of multiple factors as well as draw observations across multiple research studies.

	We call this dataset the TAWOS (\textbf{T}awosi \textbf{A}gile \textbf{W}eb-based \textbf{O}pen-\textbf{S}ource) dataset.
	It encompasses data from 13 different repositories and \numberofprojects projects, with \numberofIssues issues contributed by \numberofUsers users. 
	The dataset is publicly hosted on GitHub \cite{TAWOSonlinerepo} as a relational database, and designed such that it is amenable to future expansions by the community. Prospective contributors are welcome to join our effort to maintain, grow and further enhance the database by issuing a pull request on Github. 
		
\section{Dataset Description}
	\subsection{Data Extraction}
	This dataset was mined during the latter half of October 2020. 
	The mining process targeted 13 major open source repositories: \textit{Apache, Appcelerator, Atlassian, DNNSoftware, Hyperledger, Lsstcorp, Lyrasis DuraSpace, MongoDB, Moodle, MuleSoft, Spring, Sonatype, and Talendforge}. Most of these repositories were employed by previous work and they all used Jira as an issue management platform, which ensures uniformity of structure and availability of information.
	From each of these repositories, projects were selected such that they a) employ agile software development methodologies and b) have at least 200 issues with recorded story point entries.
	While the motivation behind the first restriction is clear, we impose the second in order to have enough data to enable statistical analyses resulting in meaningful conclusions.
	
	A total of 904 projects from the aforementioned repositories were considered, among which we selected the \numberofprojects that satisfy the collection constraints. 
	To extract issue information, we used the Jira Rest Java Client (JRJC) \cite{jrjc5.2}; 
	JRJC was used alongside our own tool, implemented in Java, to extract further features that are not implemented in JRJC (see Section \ref{sec_computed_and_derived_fields}). 
	
	\subsection{Data Storage}

\begin{figure*}
  \centering
  \includegraphics[width=\textwidth]{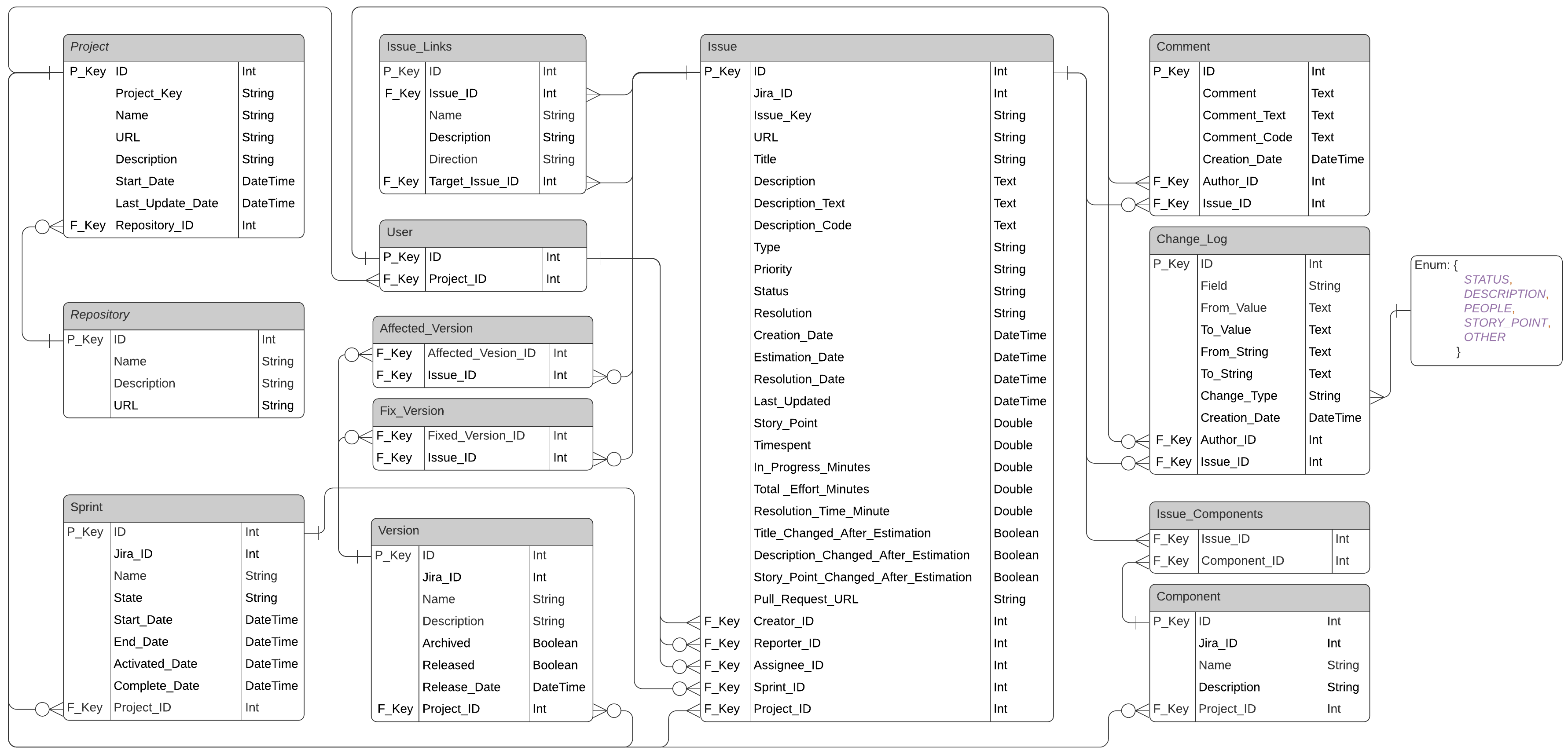}
  \caption{Entity-Relationship Diagram (ERD) for the TAWOS Issues Database.}
  \label{fig_erd}
  \Description{Entity-Relationship Diagram (ERD) for the TAWOS Issues Database.}
\end{figure*}

	The final dataset is modeled and stored as a relational database. This enables users of the dataset to employ SQL for easy horizontal and vertical data sampling in addition to allowing easier future expansion. We elected to host the dataset in the MySQL Database Management System as it is lightweight and ubiquitous.
	The database can be downloaded from a GitHub repository together with the instructions on how to install and use it \cite{TAWOSonlinerepo}.
	
	\subsection{Data Characteristics} \label{sec_data_characteristics}
		The TAWOS dataset contains \numberofIssues issues from \numberofprojects project. 
		The projects are diverse in terms of different project characteristics. 
		Each project contains issues that range from 313 to 66,741 issues. The projects span different programming languages, different application domains and different team geographical locations. 
		Table \ref{tbl_stat} shows the number of various elements for each of the projects contained in the dataset currently. Those include the number of: issues, Bug types, distinct users (i.e. bug report contributors, etc.), developers, change logs and comments, components, sprints, versions, and the number of issues with story points assigned.
		
	\subsection{Data Structure}
	Figure \ref{fig_erd} shows the Entity-Relationship Diagram of the database. 
	The core entity is the {\small \tt Issue} table, which holds the main information about an issue report. 
	Some of its fields are directly extracted from the issue report, whereas others are derived from the information stored or the events that occurred during the issue's lifecycle. We elaborate on these in Section \ref{sec_computed_and_derived_fields}.

	Other important tables are {\small \tt Comment} and {\small \tt Change\_Log} tables. 
	Comments hold the documented discussions of the team around the issue development. Change logs hold all the changes made by the users on the issue report, by recording the field that received the change, the previous value, the next value and the nature of the change. 
	Both these tables store the chronological order of the events in the {\small \tt Creation\_Date} field. 
	Information about the Sprints, Versions, and Components of the issues are also stored in separate tables. 
	The {\small \tt Issue\_Links} table captures the links between the issues. 
	The {\small \tt User} table stores all the distinct users who interacted with each project, in addition to linking the events and information to their authors and user roles. 
	Any personally identifiable information of users like their usernames and emails are redacted from this dataset.

\subsection{Computed and Derived Fields} \label{sec_computed_and_derived_fields}
To further enrich the dataset, we have augmented the mined data with several additional features that are computed or derived from the source Jira repositories as described below.

\noindent \textbf{Issue Description Text and Code}. The {\small \tt Description} field holds the long description of the user story or bug report which can contain natural text interleaved with code snippets or stack-traces. To facilitate processing, we separate the code snippets/stack traces and the natural text describing the issue into the  {\small \tt Description\_Code} and {\small \tt Description\_Text} fields respectively. We maintain the original description in the {\tt Description} field. Same is done for the {\small \tt Comment} field, from which we extract the {\tt Comment\_Code} and {\tt Comment\_Text}.
This is motivated by previous work showing that code tokens may have different meaning from those found in natural language text, hence ought to be analysed separately \cite{porru2016, scott2018, tawosi2022saner}.
		    
\noindent \textbf{Resolution Time}.
				The field {\small \tt Resolution\_Time\_Minutes} stores the time span (in minutes) between when an issue is marked for the first time as ``\textit{In Progress}'' and when it is marked as ``\textit{Resolved}''. This period can be considered as an approximation of the time taken by the development team to resolve the issue. This is usually the target variable used for bug resolution/fixing time estimation \cite{sepahvand2020predicting, lee2020continual, habayeb2017use}. Other proxies for time are provided, such as {\small \tt In\_Progress\_Time} and {\small \tt Total\_Effort\_Time}, indicating, respectively, the implementation time and the development (including code review and testing) time. 
		     
\noindent \textbf{SP Estimation Date}: 
				This field records the time when the {\small \tt Story\_Point} field of the Jira issue report was populated by the developer. This information might be useful, for example, for studies on software effort estimation, in order to properly take into account the chronological order of the estimates and avoid unrealistic usage of the data as described in previous studies \cite{jimenez2019importance, bangash2020time, sarro2020learning}.
		  
\noindent \textbf{Date and Time}.
		    The date and time stored in different Jira repositories may have different timezones, as the projects usually have contributors from all around the World. Therefore, we converted and stored all dates and times to a unified timezone, namely the Coordinated Universal Time (UTC).
		    
\noindent	\textbf{Field Change Flag.}  
		    It is important to keep track of the changes developers made to some of the issue fields. For example, the title and description of the issue are two important pieces of information used by recent automated approaches to produce effort estimates \cite{deep2018}, therefore it is important to know whether these fields have been edited after the initial estimate was done.
		    The {\small \tt Title\_Changed\_After\_Estimation} and\\ {\small \tt Description\_Changed\_After\_Estimation} fields store this flag.
		    We also provide a flag that shows whether the SP has been changed after the initial estimate.
		    Note that these flags are based on the change logs of the issue.
		    
\noindent	\textbf{Change Type in Change Log}.
		    This field is calculated to categorise change log updates into one of five categories: ``STATUS'' indicates the change is of transition type, which moves an issue from one status to another in the Jira workflow; ``DESCRIPTION'' indicates that the change was made in the issue title or description; ``PEOPLE'' indicates that the assignee or reporter of the issue
		    were changed; ``STORY\_POINT'' indicates that the Story Point field of the issue has been updated. Any other changes were categorised as ``OTHER''.
		
\bgroup
\begin{table*}
\caption{Descriptive statistics of the TAWOS dataset.}
\label{tbl_stat}
\centering
\resizebox{\textwidth}{!}{
\begin{tabular}{l | p{0.3\linewidth} | l | r r r r r r r r r r }
\toprule
Repository&Project Name&Project Key&\# Issues&\# Bugs&\# Users&\# Developers&\# Change Log&\# Comments&\# Components&\# Sprints&\# Versions&\# Story Points\\
\midrule
\hline
&Crowd&CWD&4,311&1,841&2,663&105&62,408&7,440&50&44&227&214\\
&Confluence Cloud&CONFCLOUD&23,409&10,071&24,064&513&321,439&64,655&147&477&17&352\\
&Software Cloud&JSWCLOUD&11,702&3,505&15,187&211&201,512&30,143&33&74&68&318\\
&Jira Cloud&JRACLOUD&25,669&8,339&30,020&557&295,951&74,473&66&59&170&361\\
&Confluence Server&CONFSERVER&42,324&25,477&30,755&422&1,608,633&125,591&104&565&1,121&662\\
&Atlassian Software Server&JSWSERVER&12,862&6,007&15,468&182&304,682&35,400&44&70&433&351\\
&Jira Server&JRASERVER&44,165&20,630&36,585&462&1,162,959&130,457&115&50&598&380\\
&Bamboo&BAM&14,252&6,050&7,092&107&256,321&28,638&115&14&391&528\\
&Clover&CLOV&1,501&531&347&20&25,812&2,259&15&48&63&387\\
\multirow{-9}{*}{Atlassian}&FishEye&FE&5,533&2,896&2,371&74&112,723&8,914&9&109&245&240\\
\hline
\hline
&Mesos&MESOS&10,157&4,891&1,282&252&108,349&30,152&42&227&87&3,272\\
&MXNet&MXNET&1,404&373&156&50&49,295&384&9&41&0&209\\
\multirow{-3}{*}{Apache}&Usergrid&USERGRID&1,339&349&97&37&15,435&1,535&15&38&8&487\\
\hline
\hline
&Command-Line Interface&CLI&645&399&165&29&10,956&2,233&12&98&145&374\\
&Titanium Mobile Platform&TIDOC&3,059&1,344&421&62&81,454&7,712&6&217&261&1,297\\
&Aptana Studio&APSTUD&8,135&6,152&3,365&15&107,961&19,138&49&12&91&890\\
&Appcelerator Studio&TISTUD&5,979&3,455&654&63&147,215&19,880&56&163&126&3,406\\
&The Titanium SDK&TIMOB&22,059&15,742&3,170&161&483,361&83,252&52&301&568&4,665\\
&Appcelerator Daemon&DAEMON&313&123&36&5&4,062&469&44&62&20&242\\
\multirow{-7}{*}{Appcelerator}&Alloy Framework&ALOY&1,519&646&386&30&36,312&4,491&15&118&172&315\\
\hline
\hline
DNN Tracker&DotNetNuke Platform&DNN&10,060&7,319&1,092&33&197,067&32,015&143&NA&70&2,594\\
\hline
\hline
&Blockchain Explorer&BE&802&164&149&64&8,621&1,634&0&47&0&373\\
&Fabric&FAB&13,682&3,562&1,283&457&151,811&23,056&26&142&55&636\\
&Indy Node&INDY&2,321&826&133&59&40,111&5,884&6&76&26&681\\
&Sawtooth&STL&1,663&318&174&56&15,800&576&29&22&4&966\\
\multirow{-5}{*}{Hyperledger}&Indy SDK&IS&1,531&396&177&92&21,842&2,971&10&75&30&720\\
\hline
\hline
Lsstcorp&Lsstcorp Data management&DM&26,506&2,551&277&211&310,891&71,744&259&396&4&20,664\\
\hline
\hline
Lyrasis&Lyrasis Dura Cloud&DURACLOUD&1,125&374&32&12&11,559&1,443&14&7&86&666\\
\hline
\hline
&Compass&COMPASS&1,791&737&484&17&23,617&2,077&87&91&77&499\\
&Java driver&JAVA&3,560&1,028&1,439&35&42,995&11,018&35&46&107&238\\
&C++ driver&CXX&2,032&502&409&39&30,193&4,756&13&56&70&224\\
&MongoDB Core Server&SERVER&48,663&22,342&8,837&452&1,030,545&136,823&37&NA&444&784\\
\multirow{-5}{*}{MongoDB}&Evergreen&EVG&10,299&2,636&300&67&204,228&16,939&6&NA&26&5,402\\
\hline
\hline
Moodle&Moodle&MDL&66,741&41,355&12,230&554&1,298,195&481,606&97&151&373&1,594\\
\hline
\hline
&Mule&MULE&11,816&5,421&1,449&146&233,760&16,627&129&311&274&4,170\\
\multirow{-2}{*}{Mulesoft}&Mule APIkit&APIKIT&886&467&123&34&16,137&744&19&124&96&473\\
\hline
\hline
Sonatype&Sonatype Nexus&NEXUS&9,912&5,975&2,896&82&168,909&26,159&91&143&167&1,845\\
\hline
\hline
&DataCass&DATACASS&798&166&205&10&7,070&919&11&54&154&243\\
\multirow{-2}{*}{Spring}&XD&XD&3,707&610&189&31&43,227&4,120&18&66&37&3,705\\
\hline
\hline
&Talend Data Quality&TDQ&15,315&6,288&708&131&249,243&33,438&88&144&245&1,843\\
&Talend Data Preparation&TDP&5,670&2,180&320&48&107,565&6,187&10&79&68&813\\
&Talend Data Management&TMDM&9,137&6,374&478&110&173,623&31,071&31&76&141&297\\
&Talend Big Data&TBD&4,624&2,731&553&98&70,596&5,447&35&46&149&344\\
\multirow{-5}{*}{Talendforge}&Talend Enterprise Service Bus&TESB&15,985&4,451&590&118&169,426&17,929&40&90&371&1,000\\
\hline
\midrule
&&Total&    508,963	&	237,594	&	208,811	&	6,313	&	10,023,871	&	1,612,399	&	2,232	&	5,029	&	7,885	&	69,724	\\
\bottomrule
\end{tabular}
}
\end{table*}
\egroup

\subsection{Extensibility and Maintainability}
The TAWOS database is designed such that it is easily extensible by attaching additional information to the corpus. This can help facilitate studying different problems and/or aspects of the same problem.
Sharing and managing the dataset as a GitHub repository, enables us to update, expand and enrich its content, whether by us or by the community as external contributions (i.e., pull requests). 
Github also guarantees that the information can be safely stored long-term, thus preventing the issues often faced in previous work where the data provided are not reachable anymore (e.g., due to use of volatile storing platform such as institutional webpages which change when researchers move to another institution).

\section{Originality and Relevance}
\label{sec_potential_resaerch_qs}
Previous studies have extracted information from issue reports managed in Jira to build predictive models for Story Point (SP) estimation in agile software projects \cite{porru2016, scott2018, deep2018}, however not all of them have made their data public \cite{porru2016,scott2018}. 
Choetkiertikul et al. \cite{deep2018} shared their data in a replication package \cite{deepseImpl}, however, it only consists of features considered in their study (i.e., the issue key, title, description, and story point of the mined issues).

The dataset presented herein encompasses all the projects considered in previous studies\footnote{The only exception is the MuleStudio project used by Choetkiertikul et al. \cite{deep2018}, for which we could not find the data source on-line.} \cite{porru2016, scott2018, deep2018}  augmented with more issues and features.\footnote{The TAWOS dataset has 485,650 more issues in total, and 46,411 more issues with Story Points compared to the one shared by Choetkiertikul et al. \cite{deep2018}. It also contains more issues for each of the 16 projects included in Choetkiertikul et al. \cite{deep2018}'s dataset.}. 
Furthermore, it includes 28 additional projects, which have never been used by any of these previous studies.

Our dataset has been recently used by Tawosi et al. \cite{tawosi2022deep} who analysed a total of 31,960 issues from 26 projects stored in TAWOS in order to replicate and extend the work by Choetkiertikul et al. \cite{deep2018}. 
This set of issues has also been used in a recent study on the effectiveness of clustering for SP estimation \cite{tawosi2022saner}.

We believe that the TAWOS dataset can help expedite the research in the area of Agile software development effort estimation. In addition to providing a unified benchmark for such studies, it also helps circumvent the challenges faced, and the time consumed, when mining such data from the web.
For example, we note that Choetkiertikul et al. \cite{deep2018} could not mine the same data used in the study by Porru et al. \cite{porru2016} likely because the repositories mined had changed during the time period between the two studies.

The use of different data in similar studies hinders the immensely useful opportunity to draw observations from across different studies performed at different times around a certain subject matter.
We hope that our dataset can help the community tackle this challenge. Although our dataset has been primarily designed to aid in software engineering estimation tasks, it also includes information relevant to other software engineering research, and it is designed to be expanded by other contributors. 
This allows and promotes the investigation of a wider range of SE aspects as discussed in the next section.

\section{Research Opportunities}
\label{sec_potential_resaerch_qs}
In addition to benefiting effort estimation studies, the TAWOS dataset promises value to many other areas of software engineering research, including developer productivity studies, iteration planing and task scheduling.
	
An important research topic in Requirement Engineering is \textbf{requirement prioritization} \cite{izadi2020predicting, umer2019cnn, al2016framework} and, especially in an agile setting, the selection of issues for the next iteration \cite{li2014robust, durillo2011study}.
The TAWOS dataset can support such studies by providing a large collection of issues, with known priorities and iterations (i.e., Sprints and Releases) coupled with various aspects providing a full-picture view of the issues, projects and assignees. Additionally, as the dataset makes historical project evolution from multiple repositories available, it enables cross-project analysis.
		
The TAWOS database provides information about the versioning of the software under development. 
This information includes the name, description, and release date of the version, and whether it is archived or released.
Versions connect to issues via two relations: Affected versions, and Fix versions.
The former is the version where a bug or problem was found; whereas the latter is the version where a feature is released or a bug is fixed.
This information can be used to track the bug´s lifecycle and possibly if the link to the pull request which resolves the bug is presented in the {\small \tt Pull\_Request\_URL} field, it can be tracked to the code. This information opens up avenues of research in \textbf{software testing and maintenance}.
	
The TAWOS dataset also contains information on the developer assigned to a given issue, in addition to various information regarding resolution time and the assignee's statistics. Such data enables, for example, the use of machine learning models to help automatically recommend the best developer for a new issue. Additionally, the dataset provides other useful information that can be considered for optimising \textbf{task assignment}, for example, considering developers' work load \cite{al2020workload}. 
The dataset also provides the issue status transitions, which can be used to analyse activities and events to predict the \textbf{time to fix a bug, or bug triage} \cite{sepahvand2020predicting, lee2020continual, habayeb2017use}.

\section{Final Remarks}
We have indicated just some of the research avenues the TAWOS dataset could be exploited for.
We envision that the wealth of information provided, coupled with the ability for other researchers to participate in the growth of the dataset, will enable novel research endeavours on the inter-play among several and different aspects of open-source agile software projects. 
For example, if a researcher uses our dataset to analyse the corpus of issue comments with regard to developers affects (e.g., emotions, sentiments, politeness), they can extend the dataset by issuing a pull request and thereby augmenting the existing data with the results of their investigation (e.g., augment the comments written by developers with emotions such as surprise, anger, sadness and fear). This data can be re-used in subsequent research investigating the inter-play between, for example, developer emotions and productivity.
	
We invite potential users of the database to consult our on-line documentation \cite{TAWOSonlinerepo} before use in order to understand possible limitations and select data that best fits the aim of their investigations.
We plan to curate and expand our dataset by adding other projects and features, and encourage the research community to join our effort in growing and enriching it, in order to open the door for novel research avenues.

\begin{acks}
The work of Vali Tawosi, Rebecca Moussa and Federica Sarro is supported by the ERC grant no. 741278.
\end{acks}

\bibliographystyle{ACM-Reference-Format}
\bibliography{msr2022data_arxive_v1}

\end{document}